\documentclass[twocolumns]{aa}  

\usepackage{graphicx}
\usepackage{color}
\usepackage{txfonts}
\usepackage[colorlinks=true,urlcolor=blue,citecolor=blue,linkcolor=blue]{hyperref}
\begin{document}

   \title{A novel approach to identify resonant MHD wave modes in solar pores and sunspot umbrae: $B-\omega$ analysis}
   
   \author{M. Stangalini \inst{\ref{inst1}}, D. B. Jess\inst{\ref{inst2},\ref{inst3}},
   G. Verth\inst{\ref{inst4}}, V. Fedun\inst{\ref{inst5}}, B. Fleck\inst{\ref{inst6}}, S. Jafarzadeh\inst{\ref{inst7}, \ref{inst8}}, P. H. Keys\inst{\ref{inst2}}, M. Murabito\inst{\ref{inst9}}, D. Calchetti\inst{\ref{inst10}},
   A. A. Aldhafeeri\inst{\ref{inst4},\ref{inst11}}, F. Berrilli\inst{\ref{inst10}}, D. Del Moro\inst{\ref{inst10}}, S. M. Jefferies\inst{\ref{inst12},\ref{inst13}}, J. Terradas\inst{\ref{inst14},\ref{inst15}}, R. Soler\inst{\ref{inst14},\ref{inst15}}}

   \institute{ASI, Italian Space Agency, Via del Politecnico snc, 00133, Rome, Italy\\
              \email{marco.stangalini@asi.it}\label{inst1}
              \and
              Astrophysics Research Centre, School of Mathematics and Physics, Queen’s University Belfast, Belfast, BT7 1NN, Northern Ireland, UK\label{inst2}
              \and
              Department of Physics and Astronomy, California State University Northridge, Northridge, CA 91330, USA\label{inst3}
              \and
              Plasma Dynamics Group, School of Mathematics and Statistics, The University of Sheffield, Hicks Building, Hounsfield Road, Sheffield, S3 7RH, UK \label{inst4} 
              \and
              Plasma Dynamics Group, Department of Automatic Control and Systems Engineering, The University of Sheffield, Sheffield, S1 3JD, UK\label{inst5} 
              \and
              ESA Science and Operations Department, NASA/GSFC Code 671, Greenbelt, MD 20771, USA\label{inst6}
              \and
              Rosseland Centre for Solar Physics, University of Oslo, P.O. Box 1029 Blindern, NO-0315 Oslo, Norway\label{inst7} 
              \and
              Institute of Theoretical Astrophysics, University of Oslo, P.O. Box 1029 Blindern, NO-0315 Oslo, Norway\label{inst8}
              \and
              INAF-OAR National Institute for Astrophysics, Monte Porzio Catone, RM, 00041, Italy\label{inst9}
              \and
              Department of Physics, University of Rome Tor Vergata, Via della Ricerca Scientifica 1, 00133, Rome, Italy\label{inst10}
              \and
              Mathematics and Statistic Department, Faculty of Science, King Faisal University, Al-Hassa, P.O. Box 400, Hofuf 31982, Saudi Arabia\label{inst11}
              \and
              Department of Physics and Astronomy, Georgia State University, GA 30303, USA\label{inst12}
              \and
              Institute for Astronomy, University of Hawaii, HI 96768-8288, USA\label{inst13}
              \and
              Departament de F\'isica, Universitat de les Illes Balears, E-07122 Palma de Mallorca, Spain\label{inst14}
              \and
              Institut d'Aplicacions Computacionals de Codi Comunitari (IAC$^3$), Universitat de les Illes Balears, E-07122 Palma de Mallorca, Spain\label{inst15}.}

   \date{xx-xx-xxxx}

  \abstract
   {The umbral regions of sunspots and pores in the solar photosphere are generally dominated by $3$~mHz oscillations, which are due to $p$-modes penetrating the magnetic region. In these locations, wave power is also significantly reduced with respect to the quiet Sun. However, here we study a pore where the power of the oscillations in the umbra is not only comparable, or even larger than that of the quiet Sun, but the main dominant frequency is not $3$~mHz as expected, but instead $5$~mHz. By combining Doppler velocities and spectropolarimetry and analysing the relationship between magnetic field strength and frequency, the resultant $B-\omega$ diagram reveals distinct ridges which are remarkably clear signatures of resonant MHD oscillations confined within the pore umbra. In addition to velocity oscillations, we demonstrate  that these modes are also accompanied by  magnetic oscillations, as predicted from MHD theory. The novel technique of $B-\omega$ analysis, proposed in this Letter, opens an exciting new avenue for identifying MHD wave modes in the umbral regions of both pores and sunspots.}
   
   \titlerunning{Resonant modes in a solar pore}
   \authorrunning{Stangalini M. et al.}
   \maketitle
%
%-------------------------------------------------------------------

\section{Introduction}
It is well known that large scale solar magnetic features, like pores and sunspots, are dominated by $\sim$3~mHz ($\sim$5~minute) oscillations in the photosphere and $\sim$5~mHz ($\sim$3~minute) oscillations in the chromosphere \citep[][to mention but a few]{centeno_wave_2009, centeno_spectropolarimetric_2006, 2007PASJ...59S.631N, khomenko_channeling_2008, felipe_multi-layer_2010, stangalini_mhd_2011, jess_propagating_2012, jess_influence_2013, khomenko_oscillations_2015, jess_multiwavelength_2015}. The generally accepted view is that $p$-modes are progressively absorbed by the magnetic field of sunspots and pores \citep[see for instance][]{spruit_conversion_1992, spruit_mechanisms_1996}, and converted to magneto-acoustic modes at the equipartition layer \citep{cally_umbral_1994, cally_three-dimensional_2008, cally_p-mode_2016}, which is the layer where the Alfv{\'{e}}n and sound speeds are approximately equal. These magneto-acoustic waves can propagate upwards to the chromosphere where, due to the effect of an atmospheric cutoff frequency for the magneto-acoustic waves \citep{jefferies_magnetoacoustic_2006, stangalini_mhd_2011, felipe_t_origin_2019}, the dominant period becomes $\sim$5~mHz. Interestingly, \citet{1976A&A....49..463S} reported that the locations with high oscillatory power at $\sim$5~mHz were uncorrelated with those with high oscillatory power at $\sim$3~mHz, which \citet{1986ApJ...301..992L} suggested may be a consequence of different underlying driving mechanisms. An alternative mechanism revolves around the creation of a resonance cavity in the the solar atmosphere, which has recently been shown to exist by \citet{jess_chromospheric_2020}, and independently confirmed by \citet{felipe_chromospheric_2020}. Other localized waves can be also excited due to residual convection in the umbra \citep{zhugzhda_local_2018}.

In addition to localized waves, \citet{spruit_propagation_1982}, \citet{roberts_wave_1983}, and \citet{edwin_wave_1983} predicted that magnetic flux tubes can also support the excitation of MHD global resonant modes of the structure. More recently, \citet{roberts_mhd_2019} argued that the observed velocity field in magnetic structures should therefore be regarded as the superposition of both global {\it{and}} local disturbances, giving rise to the complex oscillatory patterns generally seen in high resolution imaging of the solar atmosphere. This implies that filtering techniques are required to correctly identify the eigenmodes \citep[e.g.,][]{jess_inside_2017, albidah_proper_2020}. 

However, as of now, only low-order resonant modes have been identified, mostly in small scale magnetic structures. These include the sausage modes \citep{morton_observations_2011, martinez_gonzalez_unnoticed_2011, 2017ApJS..229....7G, keys_photospheric_2018, kang_physical_2019,2021RSPTA.37900184G}, kink modes \citep{keys_velocity_2011, stangalini_first_2013, stangalini_observational_2014, stangalini_non-linear_2015, 2017ApJS..229....9J, 2017ApJS..229...10J, keys_high-resolution_2020,2021RSPTA.37900183M}, and torsional Alfv{\'{e}}n modes \citep{jess_alfven_2009}. The majority of these works were based on the analysis of intensity or velocity fluctuations (both line-of-sight (LoS) and horizontal motions), with only a few exceptions studying the magnetic perturbations associated with, e.g., area oscillations \citep{martinez_gonzalez_unnoticed_2011, 2015ApJ...806..132G, 2016ApJ...817...44F}, or other spectropolarimetric diagnostics \citep{2021RSPTA.37900172G, 2021RSPTA.37900182K} that have recently opened the possibility of detecting true magnetic oscillations \citep{stangalini_propagating_2018}, which can help to identify clear links between such wave behavior and the chemical abundance in the solar corona \citep{2021RSPTA.37900216S, 2021ApJ...907...16B}. However, as noted above, the velocity fields inside the umbra of pores and sunspots are generally dominated by $\sim$3~mHz oscillations at photospheric heights, with only one exception reported so far to our knowledge \citep{stangalini_m_three-minute_2012}, in which the dominance of three-minute velocity oscillations at photospheric heights was reported.

Here, we further investigate the peculiar dominance of 3-minute oscillations reported by  \citet{stangalini_m_three-minute_2012} in the umbra of a magnetic pore at photospheric heights. In particular, by fully exploiting a combination of velocity and spectropolarimetric information, we are able to identify the transition of $p$-modes to a series of spectral features within the umbra that we interpret as the co-existence of multiple MHD resonant modes in the associated flux tubes.

The data set (or parts thereof)  we study here was previously used by \citet{stangalini_mhd_2011, stangalini_m_three-minute_2012}, \citet{2013A&A...560A..84S, 2016ApJ...826...49S}, and more recently by \citet{2020ApJ...890...22A}.  The study of \citet{2016ApJ...826...49S} focused on chromospheric oscillations measured in the Ca~{\sc{ii}} 854.2 nm line. They estimated the acoustic energy flux in the chromosphere and compared it to radiative losses in that region. They also reported variations of the power spectral density of intensity and velocity oscillations at chromospheric heights as a function of the distance from the center of the umbra. Here we study spectropolarimetric data obtained in the photospheric Fe~{\sc{i}} 617.3 nm line, and we extend previous analyses which considered mostly Doppler velocity oscillations by including oscillations measured in circular polarization (i.e., magnetic field, CP).  By simultaneously investigating the dependence of the power spectra of both Doppler velocity and circular polarization on the magnetic field strength we are able to identify global resonances of the magnetic structure. This allows us to detect for the first time magnetic field oscillations associated with these resonances. Compared to the power-distance plot introduced by \citet{2013A&A...560A..84S}, the new $B - \omega$ diagrams presented here offer the additional benefit that they can be applied to the study of magnetic structures with irregular shapes. Further, being based on both velocity and magnetically derived diagnostics, they fully exploit all the information that can be extracted from 2D spectropolarimetric data.

\begin{figure}[!t]
\centering 
\includegraphics[trim=150 20 100 0, width=7.8cm]{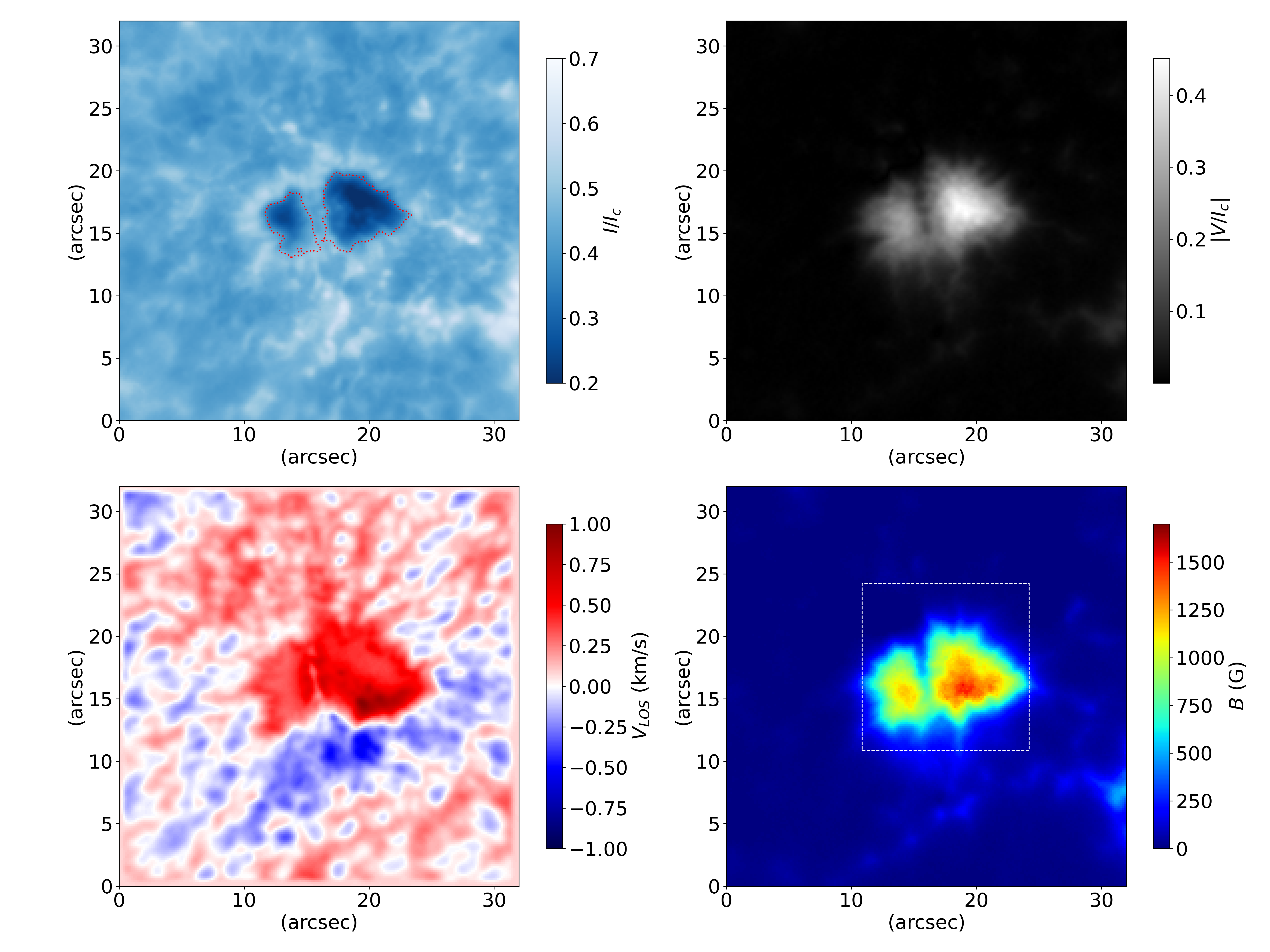}
\caption{Top left: Instantaneous Fe~{\sc{i}} $617.3$~nm line core intensity. The red dotted line represents the $B=800$ G contour used as a reference in Fig. \ref{fig:BOmega}. Top right: Magnitude of circular polarisation.  Bottom left: Instantaneous Doppler velocity map obtained in the Fe~{\sc{i}} $617.3$~nm spectral line. Bottom right: Average magnetic field strength obtained with the COG method. The white dashed box represents the region considered when filtering in $k-\omega$ space (see text and Fig.~\ref{fig:kOmega}).} 
\label{fig:maps}
\end{figure}

\section{Data set and Methods}
The data used in the work (see Fig. \ref{fig:maps}) were acquired on 2008 October 15 at 16:30~UT with IBIS \citep{cavallini_ibis_2006}, the Interferometric BIdimensional Spectrometer at the Dunn Solar Telescope (New Mexico, USA). The region observed was AR11005, which appears as a small pore with a light bridge at [25.2 N, 10.0 W].
The data set consists of 80 spectral scans in full Stokes mode, each containing 21 spectral points of the Fe~{\sc{i}} 617.3~nm line. The $\delta \lambda$ between two consecutive spectral points was 2~pm, and the exposure time was set to 80~ms. The cadence of the data is $52$~s and the pixel scale is 0.167~arcsec. Simultaneous whitelight and G-band images are restored with the multi-frame blind deconvolution \citep[MFBD;][]{van_noort_solar_2005}. Spectropolarimetric images were then co-registered and destretched to minimize the residual seeing aberrations uncorrected by the adaptive optics system. The calibration pipeline includes dark frame subtraction, flat fielding, polarimetric demodulation, and also corrects for blue-shift effects \citep{reardon_characterization_2008}.

The Doppler velocity was estimated using the method based on the estimate of the phase of the first Fourier component as in \cite{schlichenmaier_flow_2000}. The circular polarization (CP) is defined as the average of the absolute values of the Stokes-$V$ profile. In this regard, only the spectral points closest to the maxima ($10$ points) of the lobes of the profile are considered. This is done to increase the signal-to-noise ratio and exclude spectral points close to the continuum, which do not contribute much to the polarization signal, but largely to the noise contribution.

\begin{figure*}[!t]
\centering 
\includegraphics[scale=0.43]{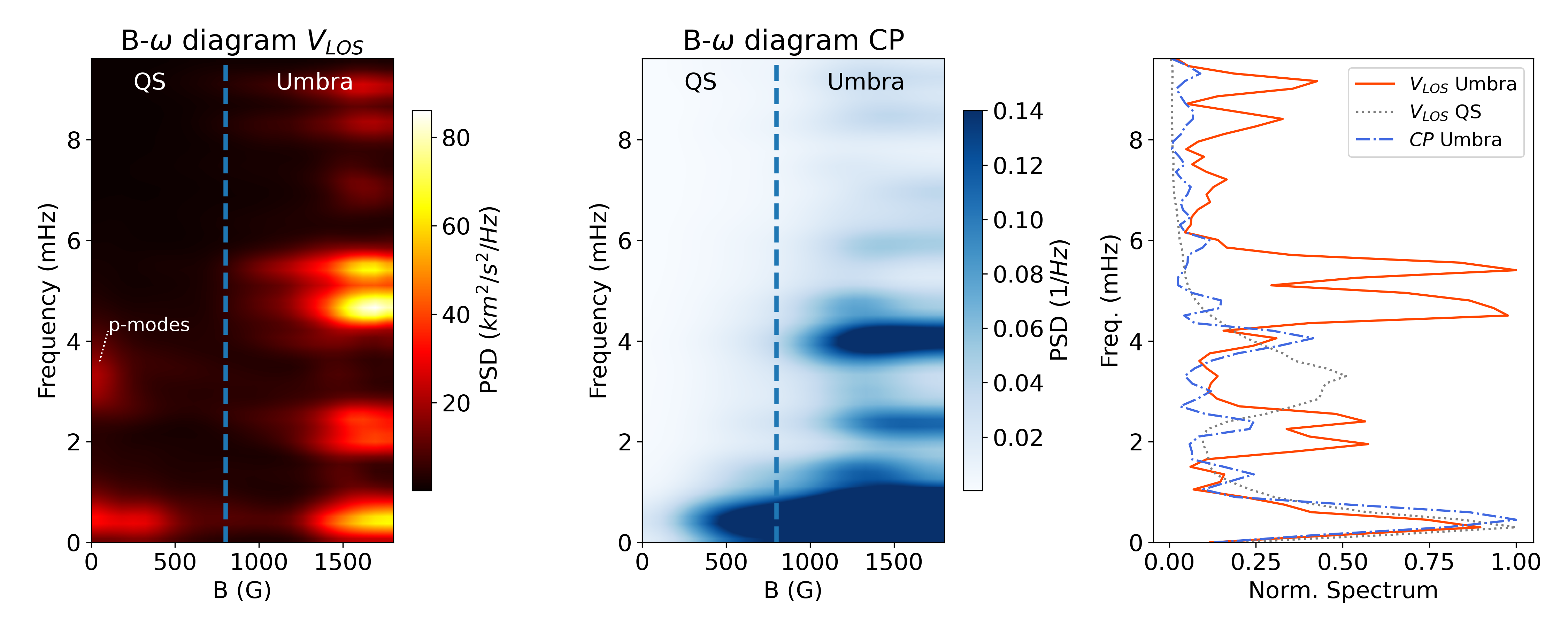}
\caption{$B-\omega$ diagram of the LoS velocity (left), CP (centre), where the vertical blue dashed line represents the approximate position of the boundary of the umbra as inferred from intensity images. Each column represents the average power spectrum across bins equal to 80~G. The global  spectra for LoS velocities and CP fluctuations, both outside and inside the magnetic structure, are shown in the right panel. These are obtained bu integrating the $B-\omega$ diagram along the horizontal axis.} 
\label{fig:BOmega}
\end{figure*}

\section{Results}
Here we investigate the spatial dependence of the dynamics on the magnetic flux. For this purpose we construct a specific diagram where each column represents the average power spectrum of pixels contained within a specific magnetic field range, which is set to an interval of 80~G in our case. This plot, which we call a $B-\omega$ diagram, and similar to the power-distance diagram investigated in \citet{2013A&A...560A..84S}, shows the modification of the power spectrum as one moves from the quiet Sun towards the inner region of the umbra (i.e., as one moves from smaller to larger values of $B$). However, the $B-\omega$ methodology can easily be applied to magnetic structures with irregular shapes, where the center of the structure itself cannot be easily and unambiguously identified. Here, $B$ is the magnetic field strength, while $\omega$ represents the oscillation frequency. This type of visualisation tool is similar to existing $k-\omega$ diagrams \citep[where $k$ represents the spatial wavenumber of the observations, e.g.,][]{duvall_frequencies_1988, krijger_dynamics_2001, rutten_dynamics_2003, kneer_acoustic_2011, jess_propagating_2012}, only now with emphasis placed on the magnitude of the embodied magnetic fields, rather than the spatial extent of their composition, thus fully exploiting the spectropolarimetric information. 

In Fig.~\ref{fig:BOmega} we show the $B-\omega$ diagram for both the LoS Doppler velocity (left) and the circular polarisation (CP; middle). Here, the CP fluctuations are a measure of the perturbations of the LoS magnetic flux. In correspondence with the umbral boundary inferred from continuum intensity maps (see Fig.~{\ref{fig:maps}}), we deduce an average magnetic field strength of $\approx$800~G (see also Fig. \ref{fig:maps} upper left panel) that segregates the umbra from the surrounding quiet Sun, which is highlighted in the left and middle panels of Fig.~{\ref{fig:BOmega}} using vertical dashed blue lines. From Fig.~{\ref{fig:BOmega}}, we observe a transition from the outside $3$~mHz oscillations corresponding to $p$-modes, to a series of spectral features inside the umbra (i.e., for $B>800$~G). It is worth noting here that none of the umbral spectral features observed in LoS velocity share the same frequency of the $p$-modes observed outside the umbra (i.e. $\sim 3$~mHz), as one would expect. 

In the right panel of Fig.~{\ref{fig:BOmega}} we plot the average spectra inside and outside the umbra to assist visualisation. Here, it is clear that frequencies corresponding to the $p$-modes (i.e., $\sim$3~mHz) are observed as a reduction of power in both LoS velocity and CP signal in the umbral locations. In turn, the umbra itself is dominated by a series of peaks at approximately $2$~mHz, $4-6$~mHz and $>8$~mHz, which are each split into  multiple sub-components. The amplitudes of these oscillations are even larger than those of the $p$-modes outside the magnetic structure. It is also worth noting that the amplitude of the oscillations in the $4-6$~mHz band is even larger than that of the surrounding $p$-modes.

From the left and middle panels of Fig.~{\ref{fig:BOmega}} it is interesting to note that these panels show different patterns of spectral features. In particular, there is no evident spectral features in the $B-\omega$ (CP) diagram corresponding to the most prominent peaks in the $B-\omega$ (LoS velocity) diagram. For example, with regard to the most prominent peak in the $B-\omega$ (CP) diagram at $\approx$4~mHz, we note that there is no equivalent increase of power in the $B-\omega$ (LoS velocity) signal. This can be better seen in the right panel of Fig.~{\ref{fig:BOmega}}, where we compare the average LoS velocity and CP spectra inside the umbra. The main spectral features present in the Doppler velocity spectrum correspond to a lack of power for CP oscillations (e.g. in the $4.5-6.0$~mHz band). In other words, there is no strict one-to-one correspondence between the LoS Doppler velocity and CP umbral spectra. However, this does not mean that corresponding to the main spectral features in LoS velocity (CP), there is absolutely no power in CP (LoS velocity).  Indeed, there exist a few spectral features dominated by CP (LoS velocity) oscillations, for which a velocity (CP) signal is also detected, although with small amplitude.

\begin{figure*}[!t]
\centering 
\includegraphics[trim=100 200 100 200, scale=0.27]{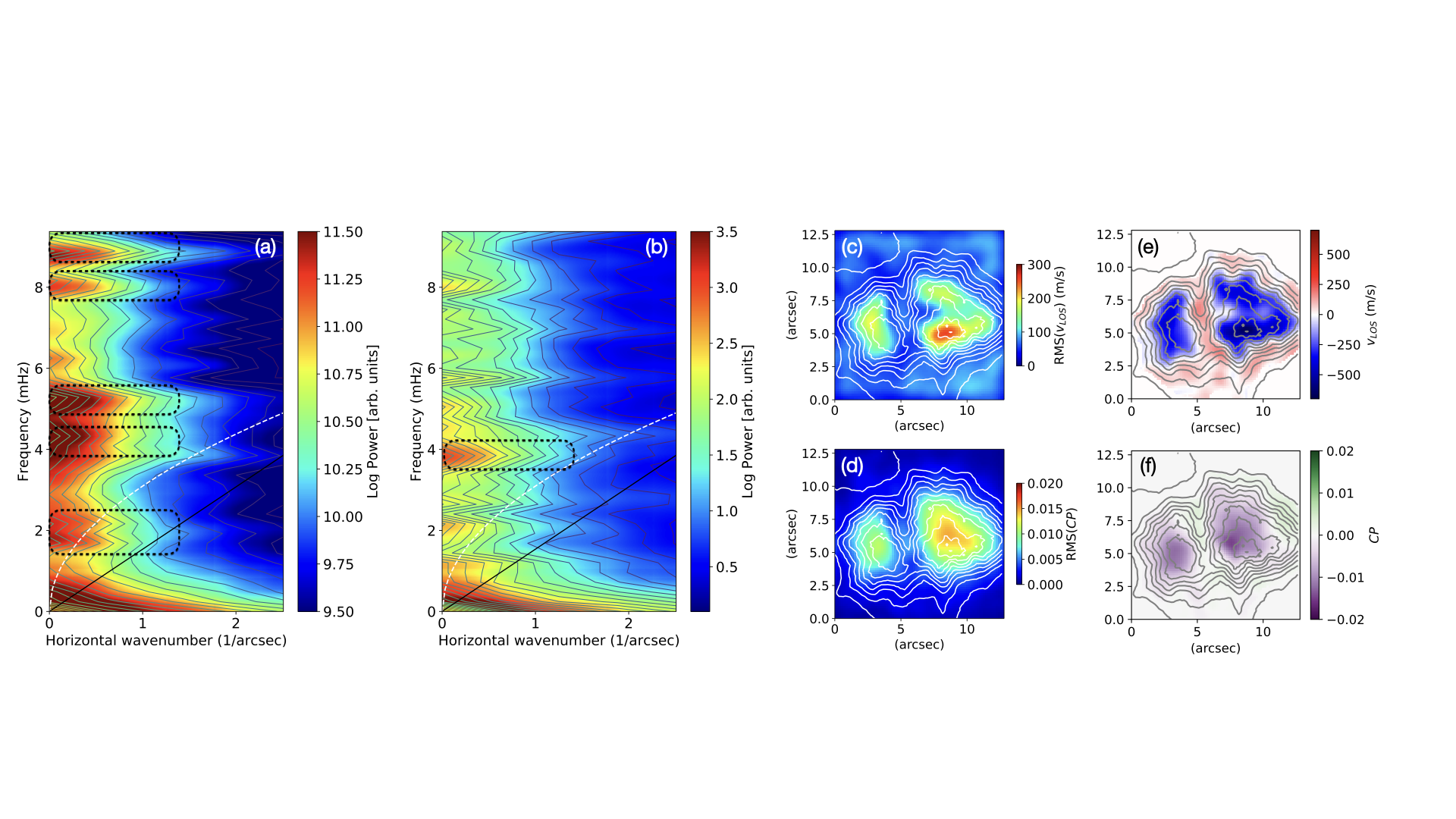}
\caption{$k-\omega$ diagrams of the LoS velocity (panel a) and CP (panel b). The white-dashed line represents the acoustic theoretical fundamental mode ($gk$, where $g$ is the gravitational acceleration and $k$ the horizontal wavenumber), while the continuous black line corresponds to the Lamb line and shows the theoretically expected propagation at the sound speed ($=c_{s}k$, where $c_{s}$ is the sound speed). Panels (c) and (d) show the root-mean-spare (RMS) amplitude of the filtered velocity and CP, respectively. Panels (e \& f) show the filtered velocities and CP signals, respectively, at one instant in time. The filter widths (in both $k$ and $\omega$ space) are illustrated by the dashed black boxes in panels (a and b), where most of the power is located.} 
\label{fig:kOmega}
\end{figure*}

\begin{figure*}
\centering 
\includegraphics[trim= 30 0 0 0, scale=0.4]{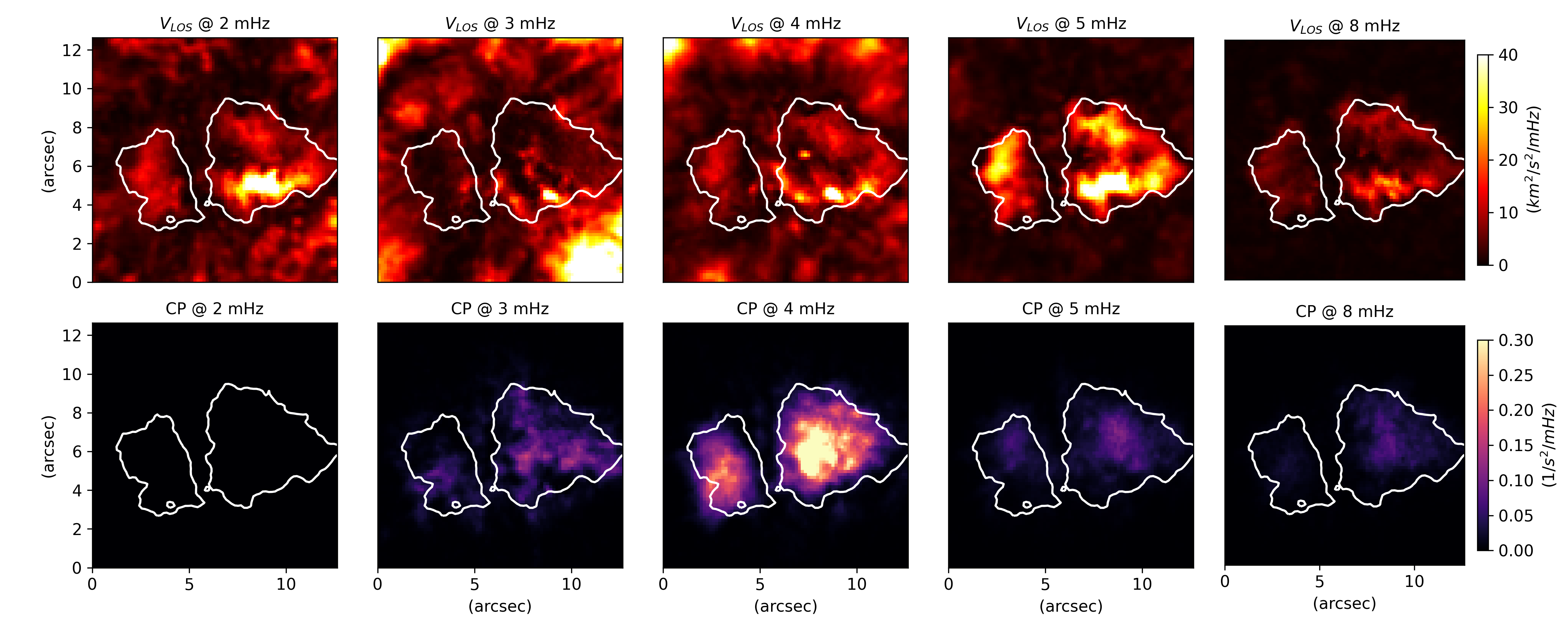}
\caption{Power maps of $v_{{\mathrm{LoS}}}$ (top) and CP (bottom) in different frequency bands ($1$ mHz width).} 
\label{fig:pmaps}
\end{figure*}

Following the approach documented by \citet{jess_inside_2017}, we produce $k-\omega$ diagrams of LoS velocity and CP, which are shown in Fig.~\ref{fig:kOmega}. Similar plots for this specific magnetic pore were also investigated by \citet{2013A&A...560A..84S}, however, here we focus on a smaller, square field-of-view (shown in the lower-right panel of Fig.~\ref{fig:maps}) to limit contamination from surrounding quiet Sun. This helps to isolate the oscillatory signals in the umbra and reliably identify a series of horizontal power enhancements (panels (a) and (b) in Fig.~{\ref{fig:kOmega}}) corresponding to the main spectral features already seen in the left and middle panels of Fig.~\ref{fig:BOmega}.
In the same figure we also show the root-mean-square (RMS) velocity and CP amplitude (panels c and d) and the instantaneous LoS Doppler velocities and CP signals after filtering in the $k-\omega$ band-pass highlighted by the dashed contours in Fig.~{\ref{fig:kOmega}}(a). Here, the filter has a Gaussian profile to limit the edge effects synonymous with discontinuities in Fourier space. These maps show the presence of radial high-order oscillations, with the oscillatory pattern largely affected by the cross-sectional shape of the magnetic structure itself. Here, we also note that, surprisingly, the CP and velocity oscillations have different spatial distributions, with the CP oscillations being more concentrated in the inner part of the umbra, while the velocity oscillations are dominant within an annulus close to the umbral perimeter. This can also be seen in the power maps at different frequencies shown in Fig. \ref{fig:pmaps}. 

\begin{figure}
\centering 
\includegraphics[trim= 30 0 0 0, scale=0.4]{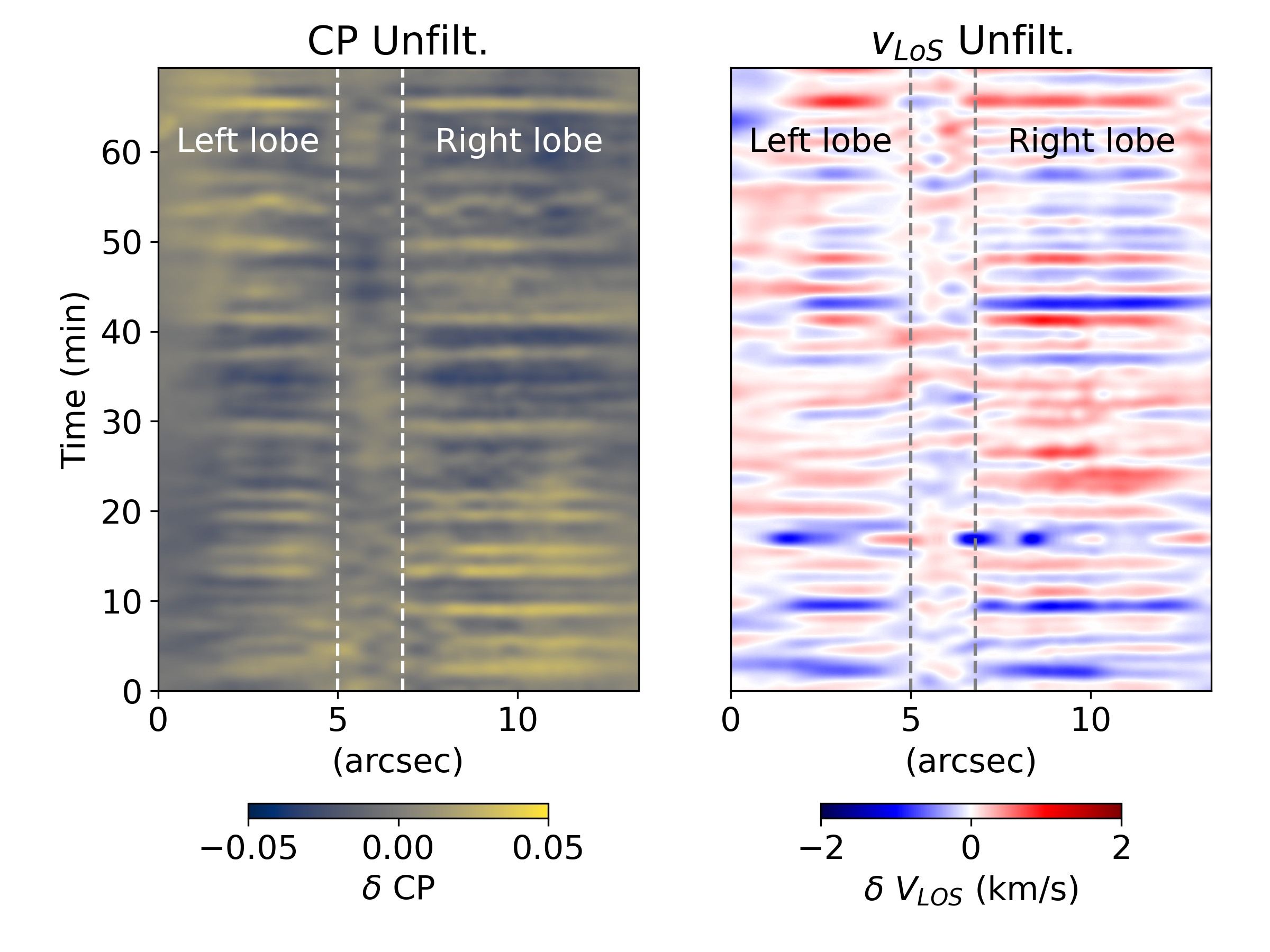}
\caption{Time-distance plots for the unfiltered CP and Doppler velocity fluctuations ($\delta {\mathrm{CP}}={\mathrm{CP}} - <{\mathrm{CP}}>$ and $\delta v_{{\mathrm{LoS}}}=v_{{\mathrm{LoS}}}-<v_{{\mathrm{LoS}}}>$) in a slice passing through the entire magnetic structure (i.e. $y=6$ arcsec with reference to left panels of Fig. \ref{fig:kOmega}). The vertical dashed lines represent the approximate position of the light bridge.} 
\label{fig:timedistance}
\end{figure}

The intensity images of the magnetic structure indicate the presence of a light bridge dividing the structure itself into two lobes. However, despite the presence of the light bridge dividing the magnetic structure into two parts, the pore behaves as a single coherent structure, at least from the perspective of the wave dynamics. This fact can be also seen in Fig. \ref{fig:timedistance}, where we plot the time-distance diagrams of the unfiltered CP and Doppler velocity fluctuations in a slice passing through the pore. The two lobes oscillate in phase, both for CP and Doppler velocity perturbations. Here, in order to highlight the perturbations with respect to the background, we have removed the temporal average before plotting each quantity.

%In Fig. \ref{fig:pmaps} we show the power maps of CP and $v_{LoS}$ for different frequency bands ($1$ mHz width). Each map is integrated over $1$ mHz. The maps are in agreement with Fig. \ref{fig:BOmega}, with CP dominated (although not only) by $4$ mHz and $v_{LoS}$ by $5$ mHz. As in the filtered maps of Fig. \ref{fig:kOmega} (panels d and e) we observe a high spatial correlation of the CP and Doppler velocity oscillatory pattern.   

\section{Discussion}
Sunspots and pores are generally dominated by $3$~mHz velocity oscillations in the solar photosphere, which are normally interpreted as the result of $p$-mode penetration into the magnetic structure, and further conversion to magneto-acoustic modes \citep[e.g.,][]{khomenko_oscillations_2015}. However, it was predicted that magnetic structures could also be affected by global resonant modes, but these have only been detected in small-scale magnetic features or in a sunspot after thorough filtering \citep{jess_multiwavelength_2015}. \citet{roberts_mhd_2019} pointed out that the complex oscillatory behaviour inside the umbra of sunspots and pores should be regarded as the superposition of both global eigenmodes of the magnetic structure and other locally excited magneto-acoustic waves and disturbances. However, the lack of dominant frequencies other than $3$~mHz in the umbra of sunspots or pores suggests that the contribution of global eigenmodes to the overall velocity field is smaller than that due to the local disturbances.

The data set explored in this work provides an ideal opportunity to study the possible presence of global resonances within a magnetic structure. Indeed, in this case, the umbra is not dominated by $3$~mHz LoS Doppler velocity fluctuations, as one would expect in the photosphere as a result of $p$-mode conversion, but by $5$~mHz oscillations that are generally only found in the chromosphere as a consequence of atmospheric stratification and the effect of the acoustic cutoff frequency \citep{felipe_t_origin_2019}.

By combining spectropolarimetric information and Doppler velocities, we have been able to characterise the wave dynamics inside the umbra and identify a series of spectral features that can be interpreted as global eigenmodes of the magnetic flux tube. These can easily be identified in a novel $B-\omega$ diagram as horizontal spectral features in LoS Doppler velocity, which arise immediately inside the perimeter boundary of the magnetic structure. This interpretation is further supported by the fact that there are no spectral features in LoS velocity at the same frequencies corresponding to $p$-modes, thus allowing for the first time the clear distinction between $p$-mode absorption and resonant modes. Furthermore, a $B-\omega$ diagram corresponding to CP signals allowed for the detection of magnetic oscillations within the umbral boundary, which are expected from MHD theory \citep{spruit_propagation_1982, roberts_wave_1983, edwin_wave_1983}.
In contrast to previous approaches based on the estimation of power as a function of distance from the center of the umbra \citep[e.g.,][]{2013A&A...560A..84S}, our approach can even be readily applied to magnetic tubes with very irregular shapes, for which the center of the structure itself cannot be easily defined. In addition, our approach does not mix the signals from magnetic and non-magnetic pixels, thus being intrinsically insensitive to contamination from ``quiet Sun'' pixels. \\
The dominant peaks in the $B-\omega$ diagram of the velocity and CP oscillations are not precisely at the same frequency. While this rules out the possibility of opacity effects and cross-talk, which would show up at the same frequency by definition, this is a rather surprising result for which we cannot offer a complete explanation yet, merely some speculative ideas. One of them is based on the detection of mixed fast- and slow-mode waves.
To illustrate this here we use the standard model made of a straight cylindrical magnetic flux tube with a purely axial magnetic field \citep[e.g.][]{spruit_motion_1981, edwin_wave_1983}. We assume a uniform magnetic field, $B_0$,  pointing in the direction of the observer, which is aligned with the $z$-direction for convenience. Under this situation velocity perturbations along the magnetic field, $v_z$, correspond to velocities along the line of sight, while magnetic perturbations along the equilibrium magnetic field, $b_z$, produce changes in CP.
Using the linearized ideal MHD equations, the amplitude ratio of the two perturbations for propagating magneto-acoustic linear waves can be written as, 
\begin{equation}
\label{eq:ampratio}
\frac{\left|\bar{b}_z\right|}{\left|\bar{v}_z\right|}=\frac{\omega_s}{\omega}\left|\frac{\omega^2}{\omega^2_s}-1\right| \ ,
\end{equation}
where $\bar{b}_z=b_z/B_0$ is the dimensionless magnetic field perturbation and $\bar{v}_z=v_z/c_s$ is the dimensionless velocity fluctuation, with $c_s$  the equilibrium sound speed inside the magnetic flux tube.

In Equation~\ref{eq:ampratio}, $\omega$ is the frequency of the propagating wave  and $\omega_s=k_z\,c_s$ is the slow (acoustic) frequency for a wave with a longitudinal wavelength equal to $2\pi/k_z$.
Under photospheric conditions, if the propagating wave is a slow MHD mode, then its frequency is typically slightly below the acoustic frequency ($\omega \lesssim \omega_s$). Therefore, from Equation~\ref{eq:ampratio} we have $\left|\bar{b}_z\right|/\left|\bar{v}_z\right|\ll 1$. Conversely, for a fast MHD wave the frequency is always above the acoustic frequency ($\omega > \omega_s$), meaning that the situation $\left|\bar{b}_z\right|/\left|\bar{v}_z\right| > 1$ is most likely to occur.
We note that in Equation~\ref{eq:ampratio} it is assumed that the waves have a propagating nature and that the line of sight is parallel to the flux tube axis. General expressions for different orientations (including standing waves) can be found in \citet{moreels_phase_2013}. In other words, under these assumptions, slow and fast MHD modes should be dominated by velocity or magnetic perturbations, respectively. While a combination of fast and slow waves is an interesting possibility to explain the presence of both Doppler velocity and magnetically dominated signals, we note that the CP $B-\omega$ diagram is dominated by a $4$ mHz peak, which is below the acoustic cutoff. Therefore, the above scenario, while providing an explanation for the high frequency power in CP, does not completely explain the dominant $4$ mHz peak. However, we should keep in mind that CP and velocity signals are extracted from different heights in the solar atmosphere. Indeed, the polarization signals originate from heights representative of the wings of the spectral line, while spectral points close to the spectral line core may contribute predominantly to the velocity signatures. 

\begin{figure}
\centering 
\includegraphics[trim= 30 20 30 30, scale=0.4]{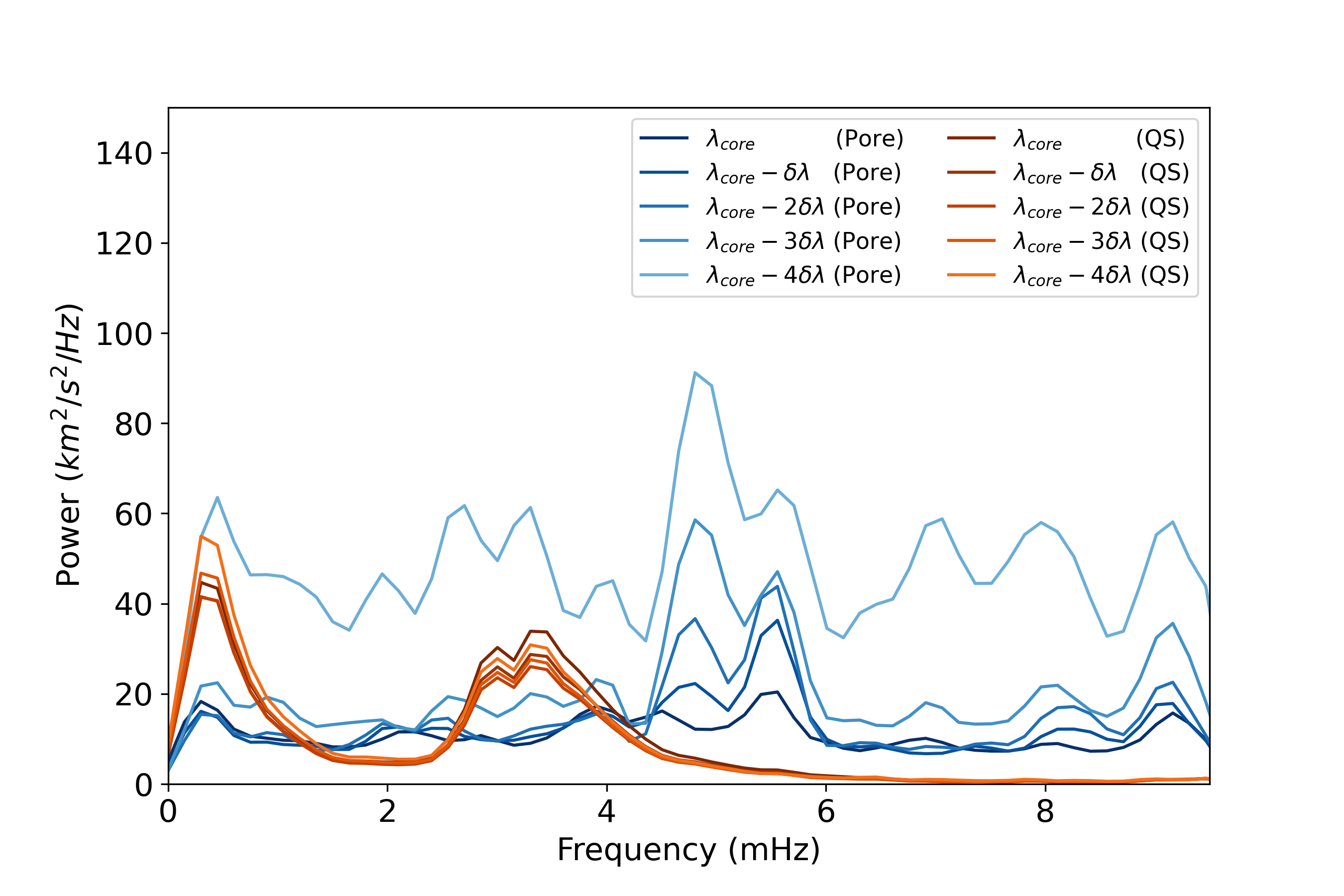}
\caption{Variation of the mean periodogram of the Doppler velocity, obtained from the bisector method from the core to the continuum, for quiet Sun (pixels below $20$ G) and the magnetic pore (pixels above $700$ G)} 
\label{fig:bisector}
\end{figure}

To illustrate this effect, in Fig.~\ref{fig:bisector} we show the power spectra of the velocity oscillations computed with the bisector method by considering different spectral positions, thus spanning different geometric heights in the solar atmosphere, from the line core forming at approximately $250$~km \citep{fleck_formation_2011}, down to the base of the photosphere corresponding to the continuum level. It is worth noting here that the line core formation height of $250$ km is representative of the quiet Sun. In magnetic structures, we would expect all the relevant formation heights to be shifted down, with the height difference between velocity and CP remaining roughly the same.  Although the range of heights explored scanning the line with the bisector method is small, the power spectra of the Doppler velocity oscillations in the magnetic umbra show large variations with geometric height; an effect not observed in the corresponding quiet Sun. In particular, heights close to the continuum show a $3$~mHz peak that is gradually suppressed as one moves upwards to the heights associated with the line core, but still the spectrum is more dominated by the peaks in the $4-6$~mHz band, alongside other higher frequency features. While these higher frequencies are also observed at the base of the photosphere (i.e. near the continuum level), they become the dominant sources of spectral power as one moves towards the spectral line core. It is worth stressing that in the quiet Sun there is essentially no difference in spectral power across the range of geometric heights. In addition, we also note that at intermediate heights, similar to those where the CP signal originate from (i.e. the wings of the line), a $4$~mHz peak is also observed. These results suggest that the method employed to measure the Doppler shifts used in this work is more sensitive to the velocities at atmospheric heights corresponding to the spectral line core. However, although at geometric heights similar to those of CP there exists a small $4$ mHz peak, the change in the formation height of the signals does not explain why in CP there is a lack of power in the $4-6$ mHz band, while this is the dominant frequency band of the Doppler velocities.

In summary, although these two options offer some interesting discussion points, in our opinion, neither of them alone or in combination can completely explain the dynamic behaviour of this magnetic pore. Although we are not able to solve this puzzle at the moment, we believe it is of paramount importance to report this surprising behavior, in the hope that state-of-the-art numerical modeling or theoretical advances will help to solve this riddle in the near future. Further, we stress that the different spatial distribution of the CP and velocity power maps (Fig. \ref{fig:pmaps}), while confirming once again the presence of a true magnetic oscillation, suggests the co-existence of two different MHD eigenmodes.

The $B-\omega$ diagrams depicted in Fig.~{\ref{fig:BOmega}} capture the transition between $p$-modes outside the umbra to global resonant modes inside it. This study therefore provides a unique exception where the two components of wave dynamics can be readily disentangled. However, a question remains as to whether this effect is visible due to the nearly complete suppression of the $p$-modes inside the umbra, or is a result of the favourable contrast between the spectral features at $3$~mHz (with respect to the others) that allows the detection of the intrinsic resonant modes of the structure. The latter appears to be the case here as the amplitude of the spectral features observed inside the umbra is even larger than that of the external $p$-modes, hence providing significant contrast for the $3$~mHz oscillations.

In our opinion, the results presented in this current study may have something to do with the excitation mechanisms of the waves and the energy associated with the external driver (e.g., turbulent convection). This suggests that the detected spectral features are associated with distinct eigenmodes, some of which show a prevalence to LoS Doppler velocity oscillations, with others to magnetic fluctuations. Nevertheless, the manifestation of magnetic oscillations suggests the presence of real MHD modes. If these structures in power are validated as resonances, this will open several new possibilities in magneto-helioseismology and might allow measuring magnetic fields in regions of the upper atmosphere that are currently difficult to probe.

\section{Conclusions}
Sunspots and pores are generally found to be dominated by $3$~mHz oscillations at photospheric heights. The consensus is that this is the result of $p$-mode penetration and absorption at the same frequency. However, it was theoretically predicted that flux tubes should also support the excitation of MHD resonant modes, which are intrinsic global oscillations of the structure. Theory also predicts that these should have an associated Doppler velocity and magnetic signal. 
A unique magnetic pore observed by IBIS at high resolution shows a dominance of $5$~mHz oscillations in the photosphere, instead of the typically expected $3$~mHz signals. By combining Doppler velocity and magnetic information obtained from polarisation measurements, is has been possible to distinguish the transition from ambient $p$-modes to internal global resonances of the magnetic structure. These are seen as a series of spectral features which arise immediately inside the umbra of the magnetic structure. Some of them are also associated with magnetic oscillations, supporting the interpretation in terms of resonant MHD modes. A novel diagnostic, in the form of a $B-\omega$ diagram that combines LoS Doppler velocities and magnetic field information at the same time clearly captures this transition, showing the progressive absorption of $p$-modes as one gets closer to the magnetic structure, followed by the onset of resonant modes inside the umbra.

\begin{acknowledgements}
The authors wish to acknowledge scientific discussions with the Waves in the Lower Solar Atmosphere (WaLSA; \href{www.WaLSA.team}{www.WaLSA.team}) team, which is supported by the Research Council of Norway (project number 262622), and The Royal Society through the award of funding to host the Theo Murphy Discussion Meeting ``High-resolution wave dynamics in the lower solar atmosphere'' (grant Hooke18b/SCTM). 
DBJ is grateful to Invest NI and Randox Laboratories Ltd. for the award of a Research \& Development Grant (059RDEN-1), in addition to the UK Science and Technology Facilities Council (STFC) grant ST/T00021X/1. 
VF, GV thank to The Royal Society, International Exchanges Scheme, collaboration with Chile (IE/170301) and Brazil
(IES/R1/191114). VF and GV are grateful to the Science and Technology Facilities Council (STFC) grant ST/V000977/1 for support provided.
SJ acknowledges support from the European Research Council under the European Union Horizon 2020 research and innovation program (grant agreement No. 682462) and from the Research Council of Norway through its Centres of Excellence scheme (project No. 262622). 
CDM would like to thank the Northern Ireland Department for the Economy for the award of a PhD studentship. 
DB is funded under STFC consolidated grant No. ST/S000240/1. 
SMJ acknowledges support under award 1829258 from the National Science Foundation.
AA acknowledges the Deanship of Scientific Research (DSR), King Faisal University, Al-Hassa, KSA for the financial support under Nasher Track (grant No.186354).
This research has received funding from the European Union’s Horizon 2020 Research and Innovation program under grant agreement No 82135 and 814335 (SOLARNET) and No 739500 (PRE-EST). 
JT and RS acknowledge the support from grant AYA2017-85465-P (MINECO/AEI/FEDER, UE).
This research has made use of the IBIS-A archive.
\end{acknowledgements}

\bibliographystyle{aa} % style aa.bst

\end{document}